\documentclass[aps,floatfix,superscriptaddress]{revtex4}
\usepackage{graphicx}
\usepackage{graphics}
\usepackage{amssymb}
\usepackage{amsmath}
\usepackage{epsfig}
\usepackage{latexsym}

\input{Qcircuit}

\newcommand{\ketb}[2]{\ket{#1}\!\!\bra{#2}}

\newcommand{\spl}{\hat{\tau}^{+}}
\newcommand{\sz}{\hat{\tau}^{z}}
\newcommand{\smin}{\hat{\tau}^{-}}
\newcommand{\id}{\hat{I}}
\newcommand{\sx}{\hat{\tau}^{x}}
\newcommand{\sy}{\hat{\tau}^{y}}
\newcommand{\da}{\downarrow}
\newcommand{\ua}{\uparrow}

\begin{document}
\title{Quantum Walks, Quantum Gates, and Quantum Computers}
\author{ Andrew P. Hines}
\affiliation{Pacific Institute of Theoretical Physics, and
Department of Physics and Astronomy, University of British Columbia,
6224 Agricultural Rd, Vancouver BC, Canada V6T 1Z1}
\affiliation{Pacific Institute for the Mathematical Sciences, 1933
West Mall, University of British Columbia, Vancouver BC, Canada V6T
1Z2}
\author{P.C.E. Stamp}
\affiliation{Pacific Institute of Theoretical Physics, and
Department of Physics and Astronomy, University of British Columbia,
6224 Agricultural Rd, Vancouver BC, Canada V6T 1Z1}

\begin{abstract}

The physics of quantum walks on graphs is formulated in Hamiltonian
language, both for simple quantum walks and for composite walks,
where extra discrete degrees of freedom live at each node of the
graph. It is shown how to map between quantum walk Hamiltonians and
Hamiltonians for qubit systems and quantum circuits; this is done for both a single-
and multi-excitation coding, and for more general mappings. Specific
examples of spin chains, as well as static and dynamic systems of
qubits, are mapped to quantum walks, and walks on hyperlattices and
hypercubes are mapped to various gate systems. We also show how to
map a quantum circuit performing the quantum Fourier transform, the key element of Shor's algorithm, to a quantum walk
system doing the same. The results herein are an essential
preliminary to a Hamiltonian formulation of quantum walks in which
coupling to a dynamic quantum environment is included.

\end{abstract}

\maketitle

\vspace{1cm}


\section{Introduction}

 \label{sec:intro}


In many quantum-mechanical systems at low energies, the Hilbert
space truncates to the point where the system is moving between a
set of discrete states (which may however be very large in number).
In this case we can describe the system, with complete generality,
as equivalent to a system in which a particle (which may itself
possess internal degrees of freedom) 'hops' between a set of
'nodes', or 'sites', on some graph - the nodes of this graph can
then be identified with states in the Hilbert space of the original
system.

The hopping amplitudes between nodes are just the transition
amplitudes in the original Hamiltonian, so that the topology of the
graph is entirely determined by these transition amplitudes. In
general we may allow the Hamiltonian to be time-dependent, so that
both the hopping amplitudes and the on-site energies are allowed to
change. We can also allow the internal state of the hopping particle
to couple to its coordinate on the graph.

In path integral language, one can think of the trajectory of a
quantum particle moving between 2 nodes A and B on this graph as a
`quantum walk', made up of a succession of discrete hops. The
amplitude to go from A to B is then given by summing over all
possible paths (or `walks') between them, with the appropriate
amplitudes.

Formulated in this way, the problem of a `quantum walk' is very
familiar to most physicists, and has in fact been under study since
the very beginning of quantum mechanics. Notable examples come from
solid-state physics (where particles hop around both crystalline
lattices \cite{crystal} and disordered systems of various
topology \cite{disorder}), from quantum magnetism \cite{Qmag} (where
an assembly of spins makes transitions between different discrete
spin states), from atomic physics and quantum optics (where one
deals with discrete atomic states, and where in the last few years
`optical lattices' have come under study \cite{atom}), and from a
large variety of problems on different sorts of graph in quantum
statistical mechanics \cite{StatM}.

\vspace{2mm}

{\it Quantum Walks and Quantum Information}: A certain
class of quantum walks has recently come under study in the
context of quantum information processing \cite{kempe03}. These walks
are intended to describe the time evolution of quantum algorithms,
including the Grover search algorithm and Shor's algorithm. The
general idea is that each graph node represents a state in the
system Hilbert space, and the system then walks in `information
space'. In some cases explicit mappings have been given between the
Hamiltonian of a quantum computer built from spin-$1/2$ `qubits' and
gates, and that for a quantum particle moving on some
graph \cite{farhi98,kempe03}. More generally, the mapping between a
walk and an algorithm is most transparent for spatial search
algorithms with the local structure of the database.

The quantum dynamics between two sites A and B on a given graph has
been shown for certain graphs to be much faster (sometimes
exponentially faster) than for a classical walk on the same graph
\cite{childs02a,shenvi03,kempe02}. It has also been argued that
quantum walks may generate new kinds of quantum algorithm, which have proved
very hard to find. Those algorithms based
on quantum walks proposed so far  fall into one of two classes \cite{ambainis03}. The
first is based on exponentially faster hitting times
\cite{farhi98,childs02,childs02a,kempe02}, where the hitting time is
defined as the mean `first passage' time taken to reach a given
target node from some initial state. While several examples have
been found, such as the `glued-trees' of Childs {\it et al.}
\cite{childs02}, there is presently no application of these to solve
some useful computational problem. The second class uses a quantum
walk search \cite{childs04,ambainis,shenvi03} providing a quadratic
speed-up. In the case of a spatial search, the quantum walk
algorithms can perform more efficiently than the usual quantum
searches based on Grover's algorithm. Amongst the graphs so far
studied for quantum walks are `decision
trees' \cite{farhi98,childs02,childs02a} and
hypercubes\cite{ambainis,childs04}; quantum walks on some other
graphs, and their connection to algorithms, were recently
reviewed\cite{kempe03}.

Several recent papers have also proposed experimental
implementations of quantum walks for quantum information processing
\cite{milburn02,fuji05}, in various systems such as ion traps,
optical lattices and optical cavities. Some of these proposals
involve walks in real space, whereas others are purely computational
walks (eg., a walk in the Hilbert space of a quantum
register\cite{fuji05}). To our knowledge, two quantum walk
experiments have been carried out: a quantum walk on the line, using
photons \cite{bouwmeester99}, and a walk on a $N=4$ length cycle,
using a 3 qubit NMR quantum computer \cite{ryan05}.  However many
experiments over the years, particularly in solid-state physics,
have also been implicitly testing features of quantum walks.

The variety of walks that one may consider is quite enormous -- one
may vary the topology of the graphs, and, as we will see below, even
quite simple walks may have a complicated Hamiltonian structure on
these graphs. Even the solid-state and statistical physics
literature has only considered a small part of the available graph
structures. In the quantum information literature, the discussion of
walks has so far been confined to a very restricted class of graphs
and Hamiltonians on these graphs. Attention has focussed almost
exclusively on either regular hypercubic lattices, on trees (or
trees connected by random links), and on `coin-tossing' walks on
lines. Often it is not obvious how one might implement these walks
in some real experiment -- clearly one is not going to be building, for example, a $d$-dimensional hyperlattice! Thus one pressing need, which
is addressed in considerable detail in the present paper, is to give
explicit mappings between the kinds of qubit or gate Hamiltonian
that one is interested in practise, and quantum walk Hamiltonians.

\vspace{2mm}

{\it Quantum Walks and Quantum Environments}: The range of possible
quantum walk systems becomes even more impressive if one notes that
any quantum walker will couple to its environment. In general one
needs to understand what form the couplings will take, and how they
will influence the dynamics of the quantum walk. Typically these
couplings can be formulated in terms of `oscillator
bath' \cite{feyV63,ajl84} or `spin bath' \cite{PS00,PCES06} models of
the environment; in the case of quantum walks we will see that
various couplings to these are allowed by the symmetries of the
problem. It has been common in the quantum information literature,
at least until very recently, to model decoherence sources and
environmental effects using simple noise sources (usually
Markovian). Results derived from such models are highly misleading --
they miss all the non-local effects in space and time which result
when a set of quantum systems are coupled to a real environment, and
also give a physically unrealistic description of how decoherence
occurs in many systems.

Thus another pressing need is to set up a Hamiltonian description of
quantum walkers coupled to the main kinds of environment which do
exist in Nature, showing how these Hamiltonians transform when one
maps between quantum walk systems and qubit or quantum gate systems.
This then allows a bridge to real experiments. This is actually a
rather substantial task which is undertaken in a separate
paper \cite{hines07}.

\vspace{2mm}

{\it Plan of paper}: The main goal of the present paper is to set up
a Hamiltonian description of quantum walk systems, and to give a
detailed derivation of the mappings that can be made between quantum
walk systems and more standard qubit and gate systems. The results
are in some cases quite complex, and in order to make them both
useful and easier to follow we give detailed results for several
examples. Two things we do not do in this paper are (i) 
incorporate couplings to the environment into the discussion - this
is the subject of another paper\cite{hines07}; and (ii)
work out the dynamics of walkers for any of the Hamiltonians we
derive (see however refs \cite{hines07,PS06}).

In section \ref{sec:QWalk-H} we begin by setting up a formalism for
the discussion of different kinds of quantum walk. In section \ref{sec:QWalk} we then
show one may systematically map from different quantum walk
Hamiltonians to various qubit systems and quantum circuits. This is
done first with single- and multi-excitation encoding of walks into many-qubit systems, and then more generally;  the mappings are
illustrated with simple examples, notably walks on a hyperlattice.
In section \ref{sec:qubit} we do the reverse, mapping qubit systems
back to quantum walks. This is done first for systems which can be
maped to spin chains, and then for more general qubit systems, both
static and dynamic; to illustrate the mappings we discuss various
chains and small qubit systems, and show how to map systems
implementing the quantum Fourier transform to quantum walks. Finally, in the concluding section \ref{sec:Conc} we summarize our
results.


\section{Quantum Walk Hamiltonians}

 \label{sec:QWalk-H}


In this section we discuss the structure of the different kinds of
quantum walk Hamiltonian we will meet. We deal in this paper with
`bare' quantum walks (ie., those without any coupling to a
background environment). We emphasize that in this section (and the
next) our primary object of study is the quantum walk, as opposed
to, eg., qubit networks or quantum circuits. However in section
4 we will be freely mapping between quantum walk systems and other
kinds of network.

We assume, as in the introduction, that the bare walk is defined by
the topology of the graph on which the system walks, and by the
`on-site' and `inter-site' terms appearing in the Hamiltonian. We
can then begin by distinguishing two kinds of bare quantum walk,
which we call `simple' and `composite', as follows:

\subsection{Simple Quantum Walk}

The `simple' quantum walker has no internal states, so that we can
describe its dynamics by a Hamiltonian with $N$ nodes, each labelled
by an integer $j\in [0,N-1]$, of form:
\begin{eqnarray}
 \label{Ham-qwa}
\hat{H}_s &=& -\sum_{ij} \Delta_{ij}(t)\left(\hat{c}_{i}^{\dagger}\hat{c}_j
+\hat{c}_{i}\hat{c}_j^{\dagger}\right) \;+\; \sum_j \epsilon_j(t)
\hat{c}_{j}^{\dagger}\hat{c}_j \nonumber \\
&\equiv& -\sum_{ij} \Delta_{ij}(t) \left(\ket{i}\!\!\bra{j}
+\ket{j}\!\!\bra{i} \right) \;+\; \sum_j
\epsilon_j(t)\ket{j}\!\!\bra{j}
\end{eqnarray}

Here each node $j$ corresponds to the quantum state $\ket{j}
=\hat{c}_j^{\dagger}\ket{0}$, so that $\ket{j}$ denotes the state
where the `particle' is located at node $j$. The two terms correspond
to a `hopping' term with amplitudes $\Delta_{ij}(t)$ between nodes,
and on-site node energies $\epsilon_j(t)$, both of which can depend
on time. There is no restriction on either the topology of the
graph, or on the time-dependence of the $\{ \Delta_{ij}(t),
\epsilon_j(t) \}$. Thus, for example, one can design a pulse
sequence for the parameters $\Delta_{ij}(t)$ and $\epsilon_j(t)$, as
a method of dynamically controlling the quantum walk.

Two of the simplest topologies that have been discussed in the
literature for quantum walks are $d$-dimensional hypercubes and
hyperlattices. The hypercube simply restricts the simple quantum
walk described above to a hypercubic graph -- its interest resides in
the fact that we can map a general Hamiltonian describing a set of
$d$ interacting qubits to a quantum walk on a $d$-dimensional
hypercube. This mapping is discussed in section \ref{sec:qubit}.
Hyperlattices extend the hypercube to an infinite lattice in $d$
dimensions; it is common to assume `translational symmetry' in the
lattice space, which means writing a very simple `band' Hamiltonian
\begin{equation}
\hat{H} = -\sum_{ij} \Delta_{o}\left(\hat{c}_{i}^{\dagger}\hat{c}_j
+ \hat{c}_{i}\hat{c}_j^{\dagger}\right) \;\; \equiv \;\; \sum_{\bf
p} \epsilon_o({\bf p}) \hat{c}_{\bf p}^{\dagger}\hat{c}_{\bf p}
 \label{H-HLo}
\end{equation}
where $\Delta_o$ is a constant, and ${\bf p}$ is the
`quasi-momentum' (also called the `crystal momentum' in the
solid-state literature); the `band energy' is then
\begin{equation}
\epsilon_o({\bf p}) = 2\Delta_o \sum_{\mu = 1}^d \cos (p_{\mu}a_o),
 \label{Eop}
\end{equation}
and the states of the walker can be defined either in the extended
or reduced Brillouin zone of quasi-momentum space. In (\ref{Eop}) we
assume a lattice spacing $a_o$, the same along each lattice vector;
(henceforth we will put $a_o = 1$). All results can be scaled
appropriately if these restrictions are lifted.

\subsection{Composite Quantum Walk}

The composite walker has `internal' degrees of freedom, which can
function in various ways. We assume these internal modes have a
finite Hilbert space, and they can often be used to modify or
control the dynamics of the walker. Thus we assume a Hamiltonian in
which the simple walker couples at each node $j$ to a mode with
Hilbert space dimension $l_j$, and on each link $\{ij\}$ between
nodes to a mode with Hilbert space dimension $m_{ij}$, and we have a
Hamiltonian
\begin{equation}
 \label{Ham-qwb}
\hat{H}_C = -\sum_{ij} \left(F_{ij}({\cal
M}_{ij};t)\hat{c}_{i}^{\dagger}\hat{c}_j  +H.c. \right) + \sum_j
G_j({\cal L}_j;t)\hat{c}_{j}^{\dagger}\hat{c}_j \;+\; \hat{H}_o(\{
{\cal M}_{ij}, {\cal L}_j \}).
\end{equation}
This composite Hamiltonian reduces to the simple walker when
$F_{ij}({\cal M}_{ij};t) \rightarrow \Delta_{ij}(t)$ and when
$G_j({\cal L}_j;t) \rightarrow \epsilon_j(t)$. We do not at this
point specify further what are $F_{ij}({\cal M}_{ij};t)$ and
$G_j({\cal L}_j;t)$, nor the form of their dynamics (which is
goverend not only by the coupling to the walker but also by their
own intrinsic Hamiltonian $\hat{H}_o(\{ {\cal M}_{ij}, {\cal L}_j
\})$), but we will study several examples below. The bulk of this
paper will be concerned with the simple walker in (\ref{Ham-qwa}),
which is already rather rich in its behaviour.

We emphasize that the internal variables are assumed to be part of
the system of interest -- that is, they are {\it not} assumed to be
part of an `environment' whose variables are uncontrolled and have
to be averaged over in any calculation. In the context of quantum
information theory these internal variables are assumed to be under
the control of the operator. For example, Feynman's original
model\cite{feynman81} of a quantum computer is a special case of a
composite quantum walk with Hamiltonian
\begin{equation}
 \label{Ham-Fqc}
\hat{H}_C = -\sum_{ij}
\left(F_{ij}(\mathbf{\tau};t)\hat{c}_{i}^{\dagger}\hat{c}_{j}  +H.c.
\right),
\end{equation}
where $\mathbf{\tau}$ corresponds to a set of register spins, where
the computation is performed. The walker implements the clock of
this autonomous computer. Another example of a composite quantum
walk is given by the Hamiltonian
\begin{equation}
\hat{H}_C = -\sum_{ij} \sum_n \delta(t-t_n) f(L_j;t)
\left(\hat{c}_{i}^{\dagger}\hat{c}_{j} +H.c. \right) \;+\;
\hat{H}_o(\{ L_j \}),
 \label{loc-C}
\end{equation}
in which decisions about where the walker hops to are made at
various times $t_l$ by discrete variables $\{ L_j \}$. Such models
include examples where some sequence of pulses acting on the
internal walker variables are used to influence its dynamics. A
simple special case of such Hamiltonians assumes the walk is
entirely on a 1-dimensional line, and that the discrete variable
$L_j$ is just a spin-$1/2$ variable -- for example, we can assume the
form
\begin{equation}
\hat{H}_C = -{1 \over 2}\sum_{j} \sum_n \delta(t-nt_o) \left[(1 +
\hat{\tau}_j^z) \hat{c}_{j+1}^{\dagger}\hat{c}_{j} \;+\; (1 -
\hat{\tau}_j^z) \hat{c}_{j-1}^{\dagger}\hat{c}_{j} \right] \;\;+\;\;
\hat{H}_o( \{ \hat{\tau}_j \}),
 \label{coinT}
\end{equation}
which is just the discrete-time coin tossing Hamiltonian, in which a
walker at site $j$ hops to the left/right depending on whether the
`coin (ie., spin-$1/2$) at this site is up/down, with decisions
being made after regular intervals of discrete time $t_o$. Obviously
one can cook up many more examples of composite walk systems.

We have sometimes found it convenient to rewrite both
(\ref{Ham-qwa}) and (\ref{Ham-qwb}) as sums over the original graph
${\cal G}$ and an ancillary graph ${\cal G}^*$ formed from the links
between the nodes of the graph. Thus we can write, for example,
\begin{equation}
\hat{H}_C = -\sum_{j' \in {\cal G}^*} \left(F_{j'}({\cal
M}_{j'};t)\hat{c}_{i}^{\dagger}\hat{c}_j  +H.c. \right) + \sum_{j
\in {\cal G}} G_j({\cal L}_j;t)\hat{c}_{j}^{\dagger}\hat{c}_j
\;\;+\;\; \hat{H}_o(\{ {\cal M}_{j'}, {\cal L}_j \})
 \label{dual}
\end{equation}
This representation puts the 'non-diagonal' or 'kinetic' terms on
the ancillary lattice on the same footing as the 'diagonal' or
'potential' terms existing on the original lattice. Such a manouevre
can be very useful in studying the dynamics of the walker, but we
will not need it in this paper.

In our study in this paper of mappings from quantum walks to systems
of qubits and/or quantum gates (or vice-versa), we will concentrate
on simple walk systems, for two reasons. First, as we will see, the
results just for simple walks are rather lengthy. Second, a proper
discussion of these mappings in a Hamiltonian framework requires a
treatment of non-local effects in time, which also arise in the
discussion of the coupling of the walker to the environment. Thus we
reserve a detailed treatment of composite walks for another paper.


\section{Encoding Quantum Walks in Multi-Qubit States}

 \label{sec:QWalk}


We would now like to map quantum walk systems to a standard quantum
computer made from qubits or quantum gates. This means that we wish
to map from a quantum walk Hamiltonian like (\ref{Ham-qwa}), acting
on states $\ket{j}$, to a qubit Hamiltonian acting on $M$ qubits;
and we require an encoding of the node state $\ket{j}$ in terms of
the $2^M$ computational basis states. We will use the following
notation for the computational basis states,
\begin{equation}
\ket{z_1z_2\ldots z_M} = \ket{z_1}\otimes\ket{z_2}\otimes \ldots\otimes\ket{z_M},
\end{equation}
where $z_k \in [\uparrow,\downarrow]$ (we use spin operators here,
instead of the more standard $[0,1]$, so as to avoid confusion with
the node indices).

We now describe two such encodings and the corresponding multi-qubit
operators needed to implement the quantum walk described by the
Hamiltonian (\ref{Ham-qwa}), thereby deriving the equivalent qubit
Hamiltonian.

\subsection{Single-excitation encoding}

Our first encoding implements the quantum walk in an $M$-dimensional subspace of the
full $2^M$ dimensional Hilbert space for $M$ qubits. In this sense,
this encoding is inefficient in its use of Hilbert space dimension.
However, the operations can prove to be more easily implementable,
requiring only two-qubit terms in the Hamiltonian.

The subspace we are interested in is spanned by the $M$-qubit states
with only a single excitation -- the states with only a single qubit
in the `up' state $\ket{\ua}_{k}$ state, with all other qubits in
the $\ket{\da}_{j}$ state (for all $j \neq k$). Each node of the
graph is then encoded via the location of the excitation (in this
case, we label the nodes from $1$ to $N$) i.e., $\ket{k} \equiv
\ket{\da}_1\otimes\ket{\da}_2
\otimes\ldots\otimes\ket{\ua}_k\otimes\ldots\otimes\ket{0}_N$. In
this encoding, the general quantum walk Hamiltonian (\ref{Ham-qwa}) is
\begin{equation}
 \label{eq::single-exc-ham}
\hat{H} = - \sum_{ i,j}\Delta_{ij}(t) \spl_{i}\smin_{j}
+\smin_{i}\spl_{j} + 2 \sum_{j} \epsilon_{j}(t)(1+\tau_{j}^z),
\end{equation}
which consists solely of $2$-qubit terms, between each connected
pair of qubits, as defined by the graph. This encoding allows the
implementation of any quantum walk using only two-qubit terms in the
Hamiltonian, provided arbitrary pairs of qubits  can interact.

To simulate evolution according to Hamiltonian
(\ref{eq::single-exc-ham}) it suffices to be able to explicitly
perform controlled evolution according to each term in the
Hamiltonian. Letting $\hat{H} = \sum_{k} \hat{H}_{k}$, for
time-idependent parameters, we use the Trotter formula
\begin{equation}\label{eq::trotter}
e^{-i\hbar t\hat{H}} \approx \left[ \prod_{k}e^{-i\hbar t H_k /N}\right]^N.
\end{equation}
approaching equality as $N \rightarrow \infty$. For time-varying
parameters in the Hamiltonian, $H(t)$, evolution is given by the
unitary
\begin{equation}
U(t,0) = \exp_+\left[-i\hbar \int_{0}^{t} H(t')dt' \right],
\end{equation}
where $\exp_+$ is the time-ordered exponential. This can be expanded
as the product
\begin{equation}
U(t,0) = U(m\delta,(m-1)\delta)\ldots U(\delta,0),
\end{equation}
for small time step $\delta = t/m$. By choosing $\delta$
sufficiently small, we approximate each term in the Hamiltonian to
be constant over this time interval,
\begin{equation}
U((n+1)\delta,n\delta) =
\exp_+\left[-i\hbar\int_{n\delta}^{(n+1)\delta}H(t')dt' \right]
\approx \exp\left[-i\hbar \delta H(n\delta)\right].
\end{equation}
Since $\delta$ is small, we then apply the Trotter formula.

So to simulate the quantum walk on a quantum  computer using this
single-excitation encoding, we must perform unitary operators of the
form
\begin{equation}\label{eq::double-term-walk}
\hat U_{ij}(\epsilon) = e^{-i\hbar\epsilon\left(\spl_{i}\smin_{j} +
\smin_{i}\spl_{j}\right)},
\end{equation}
between pairs of qubits representing connected nodes of the
corresponding graph, along with the single qubit terms
\begin{equation}\label{eq::single-term-walk}
\hat V_{k}(\epsilon) = e^{-i\hbar\epsilon \tau_z^k}.
\end{equation}
In this way, this encoding represents a `physical' walk, of a single
spin-up over a network of qubits, defined by the pairwise
interactions.

It is interesting to note the scaling of the resources required for
such a simulation of a general graph. In terms of space, the number
of qubits required for a given graph is the corresponding number of
nodes. The number of gates representing time (assuming only one- and
two-qubit operations)  is at the very least of the order of the
number of edges, assuming each qubit is in direct interaction with
all others. Details of the scaling of gate resources will depend
upon the structure of both the graph, and the quantum computing
architecture \cite{kendon03}

\subsection{Binary expansion-based encoding}

The most efficient way to encode each node is to use the binary
expansion of the integer labelling the node. We start from the state
at the `origin' of the quantum walk, and label this state by the ket
$\vert 0 \rangle$, making this equivalent to the qubit `vacuum
state' where all spins are `down'. Consider, a 2 qubit system.
Then we have the mappings $\ket{0} = \ket{\downarrow\downarrow}$,
$\ket{1} = \ket{\downarrow\uparrow}$, $\ket{2} =
\ket{\uparrow\downarrow}$, and $\ket{3} \equiv
\ket{\uparrow\uparrow}$. The number of qubits required will depend
upon the number of nodes of the graph -- $M$ qubits can encode up to
$N=2^M$ nodes. The corresponding many-qubit Hamiltonian for the
quantum walk depends upon how the nodes of the graph are labelled.
We start with the simple example of a free quantum walk on the
hypercube,  before discussing the construction for general graphs,
and quantum circuit constructions.

This encoding represents a walk in information space -- the
information about the position of the walker is stored in a quantum
register. A similar construction for the simulation of discrete-time
quantum walks on a quantum computer was conducted by Fujiwara {\it
et al.} \cite{fuji05}. Results in this section can be viewed as
analogous to this work, extended to the construction of quantum
circuits for simulating continuous-time quantum walks.

\subsubsection{Mapping a Hypercube walk to a set of qubits}

\begin{figure}[ht]
\begin{center}
\scalebox{1}{\includegraphics{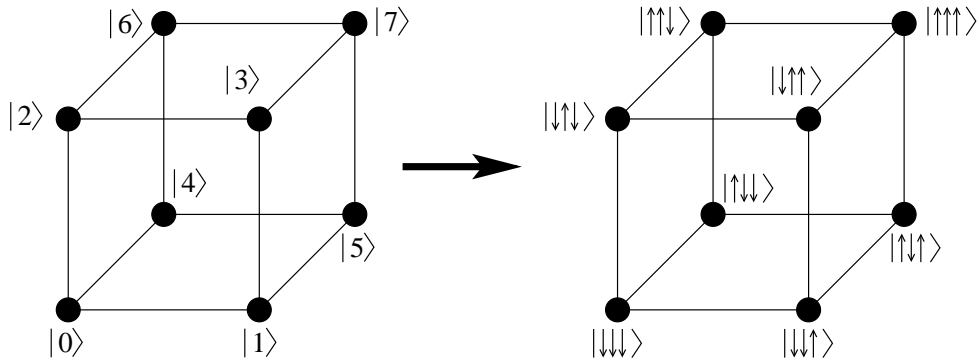}}
\end{center}
\caption{Qubit encoding of a quantum walk on the cubic lattice in
three dimensions, using three qubits.} \label{fig::cubic-lattice}
\end{figure}

Consider first the simplest possible quantum walk, where we take
$\epsilon_j = 0$ (ie., a `free walk'), and $\Delta_{ij} = \Delta_o$
in (\ref{Ham-qwa}). We also restrict the sum $\sum_{ij}$ to nearest
neighbours, so that $H = -\Delta_o \sum_{<ij>}
[\hat{c}_{i}^{\dagger}\hat{c}_j + H.c.]$. An easily visualised and
trivial example is a free quantum walk on the regular three
dimensional cube. This graph has $8$ nodes, so requires $3$ qubits
to encode. Figure \ref{fig::cubic-lattice} displays a specific
labelling \cite{kempe03} and the corresponding qubit encoding
To determine the $3$-qubit Hamiltonian corresponding to this free
quantum walk, one considers a single element, i.e.
\begin{eqnarray*}
\ketb{1}{5}& = &\ketb{\da\da\ua}{\ua\da\ua}\\
& = & \ketb{\da}{\ua}\otimes\ketb{\da}{\da}\otimes\ketb{\ua}{\ua}\\
& = & \spl\otimes\mathbb{P}_{\da}\otimes\mathbb{P}_{\ua},
\end{eqnarray*}
where $\mathbb{P}_k = \ketb{k}{k}$. Continuing this process, we obtain
\begin{eqnarray}
\hat{H} &=& -2\Delta \left( \spl\otimes\id\otimes\id +
\id\otimes\spl\otimes\id + \id\otimes\id\otimes\spl +
\textrm{H.c.}\right),\\
&=& -4\Delta\left( \sx\otimes\id\otimes\id + \id\otimes\sx\otimes\id +
\id\otimes\id\otimes\sx\right)
\end{eqnarray}
which is simply a sum of single qubit terms.

It is simple to extend this free walk to $M$-dimensions,where
$M$-qubits are required. Each qubit represents one of the $M$
orthogonal directions the quantum walker may move in from each
node,and the value of the qubit corresponding to that direction
gives at which end of that direction the walker is located. The
corresponding qubit Hamiltonian for the $M$-dimensional free quantum
walk is thus
\begin{equation}
H = -2\Delta_0 \sum_{i=1}^D \tau_{i}^x.
\end{equation}
The quantum circuit to simulate this Hamiltonian is simply single qubit rotations on each qubit,
the angle determined by the time of the walk. Scaling of resources
for the simulation is trivial -- the number of nodes $N = \log M$,
while the number of gates is the number of qubits, all of which
can be applied simultaneously.

Interactions between qubits are inevitably associated with a
`potential' $\epsilon_j$ defined over the nodes, weighted edges,
and/or next-nearest-neighbour couplings (in section \ref{sec:qubit}
below we derive the relation between the $\epsilon_j$ and
$\Delta_{ij}$ on the hypercube and the parameters of a general qubit
Hamiltonian).

\subsubsection{General walks and circuit constructions}

From the simple example of the hypercube, we can see how to
construct the multi-qubit Hamiltonian corresponding to the general
quantum walk Hamiltonian using this encoding. Each location/node is
now labeled by a bit string $\bar{z} = z_1 \ldots z_M$,
with$\uparrow \equiv 1$, $\downarrow \equiv 0$. A given on-site term
in the general quantum walk Hamiltonian (\ref{Ham-qwa}) becomes
\begin{equation}
c_{\bar{z}}^{\dagger}c_{\bar{z}} \equiv
\ket{\bar{z}}\!\!\bra{\bar{z}} = \bigotimes_{k=1}^{M}
\ket{z_k}\!\!\bra{z_k} = \bigotimes_{k=1}^{M} \mathbb{P}_{z_k} =
\prod_{k=1}^{M} \left(1-(-1)^{z_k}\hat{\tau}^z_{k}\right),
\end{equation}
where $\mathbb{P}_{z_k}$ denotes a projection operator.

For the hopping terms, we have
\begin{equation}
c_{\bar{z}}^{\dagger}c_{\bar{w}} + c_{\bar{w}}^{\dagger}c_{\bar{z}}
\equiv \ket{\bar{z}}\!\!\bra{\bar{w}}  +
\ket{\bar{w}}\!\!\bra{\bar{z}}  = \bigotimes_{k=1}^{M}
\ket{z_k}\!\!\bra{w_k} + \bigotimes_{k=1}^{M}
\ket{w_k}\!\!\bra{z_k}.
\end{equation}
For each term in the tensor product, either the bit values are
equal, and we have a projection operator, or the values are
opposite, and we have a ladder operator, ($\tau^+ , \tau^-$), such
that
\begin{eqnarray}
\ket{\bar{z}}\!\!\bra{\bar{w}}  + \ket{\bar{w}}\!\!\bra{\bar{z}} &=&  \prod_{k=1}^M (\mathbb{P}_k^{z_k})^{\delta(z_k-w_k)} \delta(1-z_k-w_k)\tau^+_k  \delta(1+z_k-w_k) \tau^-_k,\\
&=& \prod_{k=1}^M (\mathbb{P}_k^{z_k})^{\delta(z_k-w_k)} \left(\tau^x_k
+ i (z_k-w_k)\tau^y_k\right)^{1-\delta(z_k-w_k)} + \textrm{h.c.},
\end{eqnarray}
where $\delta(x)$ is the delta function. Expanding the tensor product in terms of Pauli $x$ and $y$ operators, such that the addition of the Hermitian conjugate terms ensure only products with even numbers of $\tau^y_k$ survive i.e.
\begin{equation}
\ket{\ua\da\ua\ua\ua\da}\!\!\bra{\ua\da\ua\da\da\ua} +
\ket{\ua\da\ua\da\da\ua}\!\!\bra{\ua\da\ua\ua\ua\da} =
\mathbb{P}_1^{\ua}\mathbb{P}_2^{\da} \mathbb{P}_3^{\ua}
\left(\tau^x_4 \tau^x_5 \tau^x_6 + \tau^x_4 \tau^y_5 \tau^y_6  +
\tau^y_4 \tau^x_5 \tau^y_6  - \tau^y_4 \tau^y_5 \tau^x_6 \right).
\end{equation}

To simulate the evolution of a general quantum walk on a quantum
computer using this encoding, we make use of the Trotter formula
(\ref{eq::trotter}), implying we must be able to implement unitaries
corresponding to evolution according to each term in the total
Hamiltonian. For the onsite/potential terms, this corresponds to
unitaries of the form
\begin{equation}\label{eq::onsite-U}
U(\epsilon) = e^{-i\hbar\epsilon \ket{\bar{z}}\!\!\bra{\bar{z}}}.
\end{equation}
A simple circuit to implement this unitary \cite{NC00} uses a single
ancilla qubit, initialized in the $\ket{\da}$ state, and a
multi-qubit gate which takes all qubits as input and flips the
ancilla qubit if the walker qubits are in the state $\ket{\bar{z}}$.
An example is shownbelow for the state with $\bar{z} = \ua\ua\da$,\[
\Qcircuit @C=0.75em @R=1.0em {
& & \ctrl{1} &\qw & \ctrl{1} &\qw \\
\lstick{\ket{\bar{z}}}& &\ctrl{1}&\qw &\ctrl{1}&\qw\\
& &\ctrlo{1}&\qw &\ctrlo{1}&\qw\\
\lstick{\ket{\da}} & & \targ&\gate{A_{\epsilon}}& \targ& \qw
\gategroup{1}{2}{3}{2}{0.7em}{\{} }\]
 where the solid/hollow cirlces indicate control on $\ua / \da$, and
\begin{equation}
A_{\epsilon} = \left(\begin{array}{cc} 1 & 0 \\ 0 &
e^{-i\hbar\epsilon} \end{array}\right).
\end{equation}

The multiple-controlled-NOT gates can be constructed using $3$-qubit
Toffoli gates, additional ancilla ($M-1$ gates/ancilla for
$M$control qubits)  and a controlled-NOT and  (see \cite{NC00} page
184).

For the hopping terms, we must simulate unitaries which implement
evolution according to some product of $\tau^x$'s and $\tau^y$'s on
some subset of walker qubits, if the other qubits are in some given
state -- a multi-qubit controlled operation. Firstly, the evolution
by the Hamiltonian consisting of a product over $\tau^z$ operators
can be simulated using controlled-NOT gates and a phase gate with a
single ancilla \cite{NC00}, 
\[
\Qcircuit @C=0.75em @R=1.0em {
& & \ctrl{3} &\qw & \qw & \qw & \qw & \qw &\ctrl{3} &\qw \\
\lstick{\ket{\psi}}& &\qw & \ctrl{2} & \qw &\qw &\qw & \ctrl{2}&\qw &\qw\\
& &\qw & \qw  &\ctrl{1} &\qw&\ctrl{1}&\qw&\qw&\qw\\
\lstick{\ket{\da}} & & \targ& \targ & \targ & \gate{A_{\epsilon}}&
\targ& \targ & \targ &\qw \gategroup{1}{2}{3}{2}{0.7em}{\{} }\]
which outputs $\exp\left[-i\hbar\epsilon
\tau^z_1\tau^z_2\tau^z_3\right]\ket{\psi}$. Using $U \exp[-i V] U^{\dagger} = \exp[-iUVU^{\dagger}]$ for unitaries $U$
and $V$, we can use single qubit gates and the circuit above to
simulate any product of f $\tau^x$'s and $\tau^y$'s. Since
controlled-NOT is its own inverse, the controlled evolution is
implemented by simply making the $A_{\epsilon}$ a controlled gate,
i.e.
 \[
\Qcircuit @C=0.75em @R=1.0em {
& & \qw &\qw & \qw & \qw& \ctrl{1} & \qw & \qw & \qw & \qw & \qw & \qw\\
& & \qw &\qw & \qw & \qw& \ctrlo{2} & \qw & \qw & \qw & \qw & \qw & \qw\\
& & \qw &\qw & \qw & \qw& \ctrl{4} & \qw & \qw & \qw & \qw & \qw & \qw\\
\lstick{\ket{\psi}}& & \gate{U} & \ctrl{3} &\qw & \qw & \qw & \qw &
\qw &\ctrl{3} &\qw & \gate{U^{\dagger}} &\qw \\
& & \gate{V} &\qw & \ctrl{2} & \qw &\qw &\qw & \ctrl{2}&\qw &\qw &
\gate{V^{\dagger}} &\qw\\
& & \gate{U} &\qw & \qw  &\ctrl{1} &\qw&\ctrl{1}&\qw&\qw&\qw &
\gate{U^{\dagger}} &\qw\\
\lstick{\ket{\da}} & & \qw & \targ& \targ & \targ &
\gate{A_{\epsilon}}& \targ& \targ & \targ &\qw & \qw &\qw
\gategroup{1}{2}{6}{2}{0.7em}{\{} }\] gives $\exp[-i\hbar\epsilon
\mathbb{P}^{\ua}_1 \mathbb{P}^{\da}_1 \mathbb{P}^{\ua}_1 \sx_4 \sy_5
\sx_6]$ for $U\sz U^{\dagger} = \sx$ and $V \sz V^{\dagger} = \sy$.

The complexity of the circuit to simulate a quantum walk will depend
upon the graph, and how the nodes are labelled. One simplification
is to minimize the Hamming weight (number of different bits) between
connected nodes, which we use below for the walk on the line and hyperlattice.\\

\vspace{0.2cm}

\subsubsection{ Hyperlattice walks mapped to qubits and gates}

\begin{figure}[h!]
\begin{center}
\scalebox{0.65}{\includegraphics{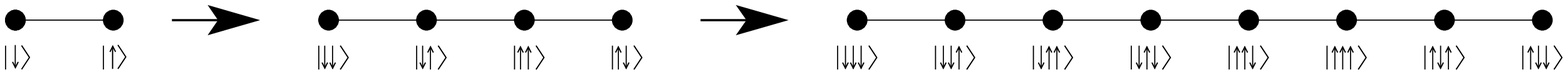}}
\end{center}
\caption{Encoding for quantum walk on the line, using $1$, $2$ and
$3$ qubits.}\label{fig::1dwalk}
\end{figure}

We start with a line with $2^N$ nodes such that the general
Hamiltonian is $H =-\sum_{i=1}^{2^N-1}\Delta_i
[\hat{c}_{i}^{\dagger} \hat{c}_{i+1}+ H.c. ] + \epsilon_i
\hat{c}_{i}^{\dagger} \hat{c}_{i} $. The encoding of the node states
is as follows: start with a single qubit, defining a two node walk,
with the nodes labelled as $\ket{\da}$ and $\ket{\ua}$. This quantum
walk is simply defined by $H = -\Delta_1 \tau_{1}^x$. Now add an
additional qubit, such that each node now has two labels, without
changing the Hamiltonian, we have two, two node walks, which we now
join together at opposite ends, such that the order of the nodes is
now $\da\da$,$\da\ua$,$\ua\ua$,and $\ua\da$. We then continue in
this fashion  (as shown in the figure \ref{fig::1dwalk}) for N-qubits,
giving a $2^N$ node walk on the line. Note that the label of each
node differs from it's nearest neighbours in only one bit.
Given a bit-string $\bar{x} = x_N x_{N-1} \ldots x_2x_1 $ specifying
a node, the position along the line (with $\da\da\ldots \da$
corresponding to the origin, ie., position $1$) is given by the
function
\begin{equation}
F(\bar{x}) = 1 + \sum_{n=1}^{N} 2^{N-n}\left(\bigoplus_{i=0}^{n-1}
x_{N-i} \right),
\end{equation}
where $\oplus$ denotes addition modulo $2$.

This labelling results in the following $N$-qubit Hamiltonian for
the quantum walk on the line,
\begin{equation}
\sum_{m=1}^N \left(\sum_{\bar{x}: F(\bar{x}) = 2^{m-1}(2n+1),
n=0,1,\ldots} \Delta_{F(\bar{x})} \spl_{m} \prod_{n \neq m}
\mathbb{P}^{x_n}_n + \textrm{H.c.}\right) + \sum_{\bar{x}}
\epsilon_{F(\bar{x})} \ket{\bar{x}}\!\!\bra{\bar{x}},
\end{equation}
such that each hopping term consists of only one Pauli term, and the
rest projection operators. For the corresponding circuit simulation,
this means that only multiply-controlled \emph{single}-qubit gates
are required. In the case of uniform hopping, $\Delta_i = \Delta_0$,
the sum over the hopping terms simplifies to
\begin{equation}
H_{hop} = -2\left( \sx_{1} + \sx_{2} \mathbb{P}_{\ua}^{(1)} +
\sx_{3} \mathbb{P}_{\ua}^{(2)}\mathbb{P}_{\da}^{(1)} +
\sx_{4}\mathbb{P}_{\ua}^{(3)}\mathbb{P}_{\da}^{(2)}\mathbb{P}_{\da}^{(1)}+
\ldots +\sx_{N} \mathbb{P}_{\ua}^{(N-1)}
\mathbb{P}_{\da}^{(N-2)}\ldots \mathbb{P}_{\da}^{(1)} \right)
\end{equation}
The corresponding circuit to simulate $U_k(\epsilon) = \exp{(-i\hbar
\epsilon H_k)}$, for $\epsilon = t/N$ with $H_k =
\sx_{k}\mathbb{P}_{\ua}^{(k-1)}\mathbb{P}_{\da}^{(k-2)}\ldots
\mathbb{P}_{\da}^{(1)}$, (such that the corresponding unitaries
$U_k$ are controlled rotations on the $k^{th}$ qubit.) is shown
below (for $6$ qubits);
\[
\Qcircuit @C=0.75em @R=1.0em {
& \gate{X_{4\epsilon}} & \ctrl{1} & \gate{X_{\pi}} &\ctrl{1}&\qw&\ctrl{1}&
\qw & \ctrl{1}& \qw &\ctrl{1}& \gate{X_{\pi}}& \qw\\
& \qw & \gate{X_{4\epsilon}} & \qw & \ctrl{1} & \gate{X_{\pi}} &\ctrl{1}&\qw
&\ctrl{1} & \qw & \ctrl{1}& \gate{X_{\pi}}&\qw\\
& \qw &\qw & \qw&\gate{X_{4\epsilon}} & \qw & \ctrl{1}&
\gate{X_{\pi}}&\ctrl{1} &\qw & \ctrl{1}& \gate{X_{\pi}}&\qw \\
& \qw &\qw & \qw & \qw & \qw&\gate{X_{4\epsilon}}& \qw &
\ctrl{1} &\gate{X_{\pi}} &\ctrl{1} & \gate{X_{\pi}}&\qw \\
& \qw &\qw & \qw & \qw & \qw& \qw & \qw & \gate{X_{4\epsilon}}
& \qw & \ctrl{1} & \qw &\qw \\
& \qw &\qw & \qw & \qw & \qw &\qw & \qw &\qw&\qw & \gate{X_{4\epsilon}}
& \qw &\qw}
\]\label{fig::qwcirc-1}
where we have used the notation $X_{\theta} \equiv R_x({\theta}) = \exp(-i\hbar\theta \sx/2)$, such that  $X_{\pi}$
gate corresponds to the Pauli-$X$ i.e. a bit flip.

To write the circuit above in terms of a one- and two-qubit gates
we use the construction described above. Explicitly, we require the
multiply controlled gate
\[
\Qcircuit @C=0.75em @R=1.5em {
& & & & & & & \lstick{\ket{c_1}}& \ctrl{1}& \qw & \qw & \qw& \qw &
\qw & \qw& \qw& \qw& \qw &\ctrl{1} & \qw\\
& & & & & & & \lstick{\ket{c_2}}& \ctrl{5}& \qw & \qw & \qw& \qw &
\qw& \qw& \qw& \qw& \qw &\ctrl{5} & \qw\\
& & & & & & & \lstick{\ket{c_3}}& \qw & \ctrl{4} & \qw & \qw &
\qw & \qw & \qw& \qw& \qw& \ctrl{4} &\qw & \qw\\
\lstick{\ket{c_1}}& \ctrl{1}& \qw & & & &  &\lstick{\ket{c_4}}&
\qw& \qw & \ctrl{4} & \qw &\qw & \qw& \qw& \qw&\ctrl{4} & \qw&
\qw & \qw\\
\lstick{\ket{c_2}}& \ctrl{1}& \qw & & &  & &\lstick{\ket{c_5}}&
\qw & \qw & \qw & \ctrl{4} &\qw & \qw & \qw& \ctrl{4} & \qw&
\qw & \qw & \qw\\
\lstick{\ket{c_3}}& \ctrl{1}& \qw & & &  & &\lstick{\ket{c_6}}&
\qw & \qw & \qw & \qw & \ctrl{4} &\qw  & \ctrl{4}& \qw& \qw &
\qw & \qw & \qw\\
\lstick{\ket{c_4}}& \ctrl{1}& \qw & & \lstick{\equiv}& & &
\lstick{\ket{0}}& \targ& \ctrl{1}&\qw& \qw & \qw &\qw & \qw&
\qw& \qw& \ctrl{1}&\targ & \qw\\
\lstick{\ket{c_5}}& \ctrl{1}& \qw & & & & &\lstick{\ket{0}}&
\qw& \targ & \ctrl{1} & \qw & \qw & \qw & \qw & \qw &\ctrl{1}
& \targ & \qw & \qw\\
\lstick{\ket{c_6}}& \ctrl{1}& \qw & & & & &\lstick{\ket{0}}&
\qw & \qw & \targ &\ctrl{1} & \qw & \qw& \qw &\ctrl{1} &
\targ& \qw & \qw & \qw\\
\lstick{\ket{tg}}&  \gate{R_x(2\epsilon)}& \qw & & & & &\lstick{\ket{0}}
& \qw & \qw & \qw & \targ & \ctrl{1} & \qw &\ctrl{1}& \targ &
\qw& \qw& \qw & \qw\\
& & & & & & &\lstick{\ket{0}}& \qw & \qw &\qw &\qw &\targ &
\ctrl{1} & \targ& \qw& \qw& \qw& \qw & \qw\\
& & & & & & &\lstick{\ket{tg}}& \qw & \qw & \qw & \qw & \qw
&\gate{R_x(2\epsilon)} & \qw& \qw& \qw& \qw& \qw & \qw }
\]
with the Toffoli gates  realised using single qubit rotations
and CNOT gates, as shown below:
\[
\Qcircuit @C=0.75em @R=1.0em {
&\ctrl{1} & \qw& & & &\qw &\qw&\qw & \qw &\qw & \ctrl{1} &\qw
&\qw &\qw &\qw &\qw & \ctrl{1} &\qw & \ctrl{2}&\qw  & \ctrl{2}
&\qw  &\qw &\qw  \\
&\ctrl{1}& \qw & & \equiv& &\qw & \ctrl{1} &\qw &\ctrl{1} &
\gate{\hat{\alpha}} & \targ &\qw  & \ctrl{1} &\qw  &\ctrl{1}
&\gate{-\hat{\alpha}} & \targ &\qw &\qw &\qw &\qw &\qw &\qw &\qw \\
&\targ &\qw& & & & \gate{C} &\targ & \gate{B} &\targ & \gate{A}
& \qw& \gate{C} &\targ & \gate{B^{-1}} &\targ & \gate{A'} &
\qw & \gate{C} &\targ & \gate{B} &\targ & \gate{A} &\qw
 }
\]
using the following single qubit gates (where $R_{a}(\theta) =
\exp(-i\theta\sigma_a / 2)$):
\begin{eqnarray*}
A = R_z(\frac{-\pi}{2})R_y(\frac{\pi}{4}), &\,\, B =
R_y(\frac{-\pi}{4}),& \,\, C = R_y(\frac{\pi}{2}) \\
A' = R_z(\frac{-\pi}{2})R_y(\frac{-\pi}{4}), &\,\,
\hat{\alpha} = \left[ \begin{array}{cc} 1 & 0 \\ 0 & e^{-i\pi/4}
\end{array}\right], &\,\, -\hat{\alpha} =
\left[ \begin{array}{cc} 1
& 0 \\ 0 & e^{i\pi/4} \end{array}\right]
\end{eqnarray*}

Finally, we need to be able to apply a controlled-$R_x(2\epsilon)$,
which is simply:
\[
\Qcircuit @C=0.75em @R=1.0em {
&\ctrl{2}& \qw & & & &\qw & \ctrl{2} &\qw &\ctrl{2} & \qw &\qw \\
& & & & \equiv & & & & & & &\\
&\gate{X_{2\epsilon}} &\qw& & & &\gate{F} &\targ & \gate{E} &\targ
& \gate{D} &\qw }
\]
where
\begin{eqnarray*}
D = R_z(\frac{-\pi}{2})R_y(\frac{\epsilon}{2}), &\,\, E =
R_y(\frac{-\epsilon}{2}),& \,\, F = R_z(\frac{\pi}{2}).
\end{eqnarray*}\\
This can be simply modified to the quantum walk on the circle, by
modifying the last term in the Hamiltonian to $
\sx{N}\mathbb{P}^{(N-2)}_{\da} \ldots \mathbb{P}^{(0)}_{\da}$,
and in turn altering the corresponding gate. Having the hopping
amplitudes between nodes equal greatly simplifies the quantum
circuit simulation -- the number of gates requires
scales approximately as $\mathcal{O} (n^2)$ for each incremental
time step.

The construction of the qubit quantum circuit for simulating the
quantum walk in the line can be easily generalised to simulate a
quantum walk on an arbitrary $D$-dimensional hyperlattice, with $2^{ND}$ nodes.

Each node on the hyperlattice is specified by $D$ bit-strings
of length $N$, each of which denote the location of the node
in a given
direction -- each node is represented by an $N \times D$ qubit
state, $\ket{\bar{x}_1;\bar{x}_2;\ldots;\bar{x}_D}$, where
$\bar{x}_k$ is an $N$-bit string.
\begin{figure}[h!]
\begin{center}
\scalebox{0.45}{\includegraphics{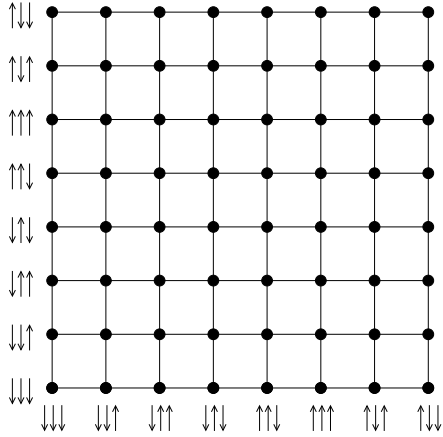}}
\end{center}
\caption{Encoding for quantum walk on the two dimensional lattice.
Each node is encoded via two bit strings, of length $3$ in this
case.}
 \label{fig::2dwalk}
\end{figure}

Using this encoding, the quantum walk on the hyperlattice simply
corresponds to $D$ individual quantum walks on the line, where $D$
is the dimension of the lattice -- there is no interaction between
qubits specifying different directions. Thus, we use the above
construction on $D$ different sets of $M$-qubits to define the
quantum walk on the $D$-dimensional hyperlattice as follows
\begin{equation}
H = \sum_{d=1}^{D} \left[ \sum_{m=1}^N \left(\sum_{\bar{x_d}:
F(\bar{x_d}) = 2^{m-1}(2n+1), n=0,1,\ldots} \Delta_{F(\bar{x_d})}
\spl_{m_d} \prod_{n_d \neq m_d} \mathbb{P}^{x_{n_d}}_n +
\textrm{H.c.}\right) + \sum_{\bar{x_d}} \epsilon_{F(\bar{x_d})}
\ket{\bar{x_d}}\!\!\bra{\bar{x_d}}\right].
\end{equation}
\vspace{.5cm}

We have discussed the construction of qubit Hamiltonians for a given
walk when the graph structure is completely known. Another scenario
is where we are given access to a `black-box' or oracle, which contains
information about the graph structure, e.g the adjacency matrix. In
the standard set-up, we may query the oracle with two nodes to
determine if there is such a connection. This is the situation in
the Childs {\it et al.} algorithm \cite{childs02}, and was
considered more generally by Kendon \cite{kendon03}.


\section{From Qubit Hamiltonians to Quantum Walks}
\label{sec:qubit}


The other direction to approach these mappings from is to start with a multi-qubit
Hamiltonian, and determine a corresponding quantum walk. We begin with a simple one-dimensional spin chain.

\subsection{Spin-chains to Quantum Walks}
 \label{sec:spinC}

The $XY$ model in one dimension corresponds to a chain of $N$ qubits
(spin-$\frac{1}{2}$ particles) with nearest-neighbour couplings,
described by the Hamiltonian
\begin{equation}\label{eq::Ham-XY}
\hat{H}_{XY} =  \sum_{i=1}^{N} -\frac{J}{2}\left(\sx_{i}\sx_{i+1}
+\sy_{i}\sy_{i+1}\right) + \frac{h}{2} \sz_{i},
\end{equation}
which assumes homogenous coupling strengths, $J$. This model is
exactly solvable using the Jordan-Wigner transformation, mapping the
model to a system of spinless fermions. In this representation, the
Hamiltonian has a natural quantum walk interpretation, as fermions
hopping between sites. The Jordan-Wigner transformation defines the
fermionic operators
\begin{equation}
\hat{c}_i  =  \left(\prod_{j<i} \sz_{j} \right) \spl_{i},
\;\;\;\;\;\;\;\;\;\;\;\;\; \hat{c}^{\dagger}_i  = \left(\prod_{j<i}
\sz_{j} \right)\smin_{i},
\end{equation}
which respect the fermionic canonical commutation relations,
$\{\hat{c}_i,\hat{c}_j^{\dagger} \} = \delta_{ij}$ and
$\{\hat{c}_i,\hat{c}_j \} =
\{\hat{c}_i^{\dagger},\hat{c}_j^{\dagger} \} = 0$. The spin
operators are expressed as
\begin{eqnarray}
\sz_{i} & = & \hat{I} - 2\hat{c}_i^{\dagger}\hat{c}_{i},\\
\spl_{i} & = & \prod_{j<i} (\hat{I} - 2\hat{c}_{j}^{\dagger}\hat{c}_{j}) \hat{c}_i,\\
\smin_{i} & = & \prod_{j<i} (\hat{I} - 2\hat{c}_{j}^{\dagger}\hat{c}_{j}) \hat{c}_i^{\dagger}.
\end{eqnarray}
The $XY$ Hamiltonian then becomes
\begin{equation}
\hat{H}_{XY} = \frac{h}{2} + \sum_{i=1}^{N}
-J\left(\hat{c}_{i+1}^{\dagger}\hat{c}_i +
\hat{c}_{i}^{\dagger}\hat{c}_{i+1} \right) -
h\hat{c}_{i}^{\dagger}\hat{c}_i
\end{equation}
describing free, spinless fermions, hopping along a $1$-dimensional
lattice, since the total fermion number, $\hat{n} = \sum_{i=1}^{N}
\hat{c}_{i}^{\dagger}\hat{c}_i$, is conserved.

\begin{figure}[ht]
\begin{center}
\scalebox{0.7}{\includegraphics{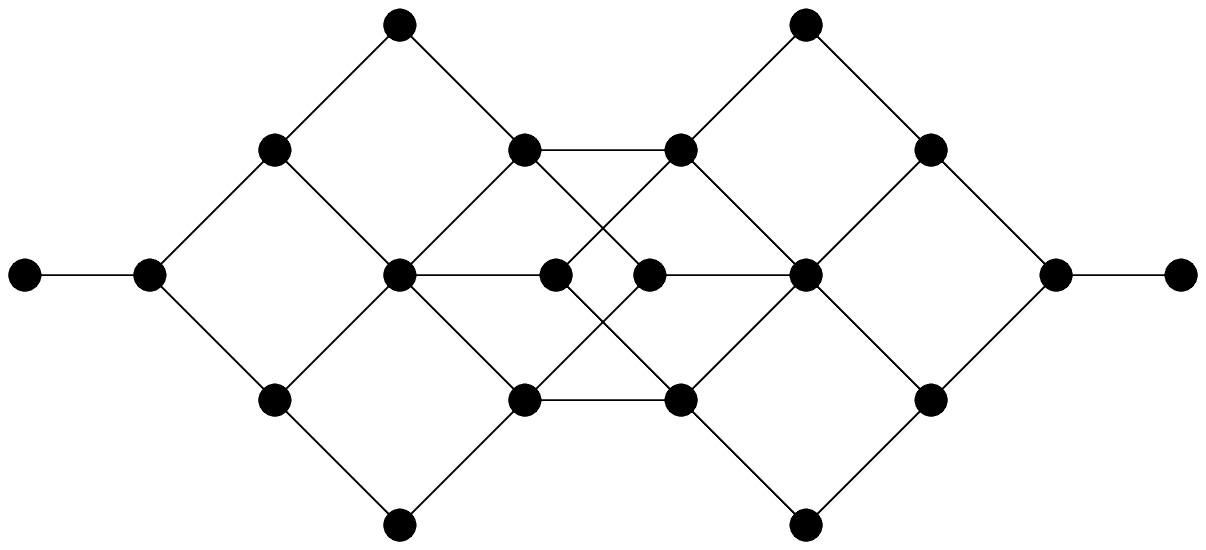}}
\end{center}
\caption{Graph for the quantum walk given by the Hamiltonian,
$\hat{H}_{XY}$, with $6$ sites, and three excitations.}
\label{fig::example-higherD}
\end{figure}
It is interesting to consider the same system with a higher number
of excitations. In this case, the dynamics is restricted to a
subspace with dimension $D = \left(\begin{array}{c} N \\ n
\end{array}\right) = \frac{N!}{n!(N-n)!}$, where $n$ is the number
of fermions/excitations. Now consider each state as encoding a
node of a graph, reverting to the binary-encoding. The symmetry in
the system results in interesting graphs for the corresponding
quantum walk. For example, the $N=6, n = 3$ case, where nodes are
encoded by states of the form
$\ket{\uparrow\uparrow\uparrow\downarrow\downarrow\downarrow}$ with
three spins up, and three down, is shown in figure
\ref{fig::example-higherD}, where the two end states correspond to
$\ket{\uparrow\uparrow\uparrow\downarrow\downarrow\downarrow}$ and
$\ket{\downarrow\downarrow\downarrow\uparrow\uparrow\uparrow}$. We
see that this graph has a tree-like structure, leading into a cube
in the middle. The continuous quantum walk on this graph is exactly
solvable.

\begin{figure}[ht]
\begin{center}
\scalebox{0.9}{\includegraphics{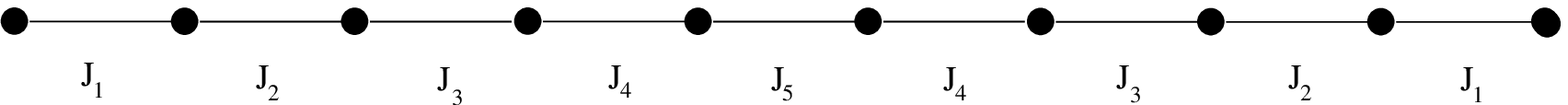}}
\end{center}
\caption{Graph for the quantum walk given by the Hamiltonian,
$\hat{H}_{XY}$, with $6$ sites, and three excitations, reduced to a
linear chain. The couplings are $J_1 =1,\, J_2 = \sqrt{2},\, J_3 =
4/\sqrt{6},\, J_4 = 5/3$ and $J_5 = 2$.}
\label{fig::corresp-lin-chain}
\end{figure}

It is possible to `collapse' such a quantum walk to a biased walk
along a line \cite{childs02a}. This corresponds to the $XY$-model
with non-homogenous coupling strengths. This is done be defining
column subspaces, such that states in column space $k$, are only
connected to states in column spaces $k-1$ and $k+1$, in terms of
the corresponding graph for the quantum walk. Site $k$ on the line
then corresponds to an equal superposition of states in the
corresponding column subspace. The strength of the coupling between
the nodes is then determined from the Hamiltonian. Figure
\ref{fig::corresp-lin-chain} shows the linear chain corresponding to
to the $XY$-Hamiltonian with six sites, in the three excitation
subspace. The two end nodes correspond to the states
$\ket{\ua\ua\ua\da\da\da}$ and $\ket{\da\da\da\ua\ua\ua}$.

\subsection{Static Qubit Hamiltonians to Quantum Walks}
 \label{sec:QB-QW}

Now let's look at more general spin systems. A system of
considerable interest, both methodological and practical, is the
general $N$-qubit Hamiltonian with time-independent couplings. As an
example consider the following form:
\begin{equation}\label{eq::HamUQC}
\hat{H} = \sum_{n=1}^{N} \left(\epsilon_n \sz_{n} + \Delta_n
\sx_{n}\right) - \sum_{i,j} \chi_{ij} \sz_{i}\sx_{j} +
\sum_{i<j}V_{ij}^{\perp} \sx_{i}\sx_{j} +  V^{\parallel}_{ij} \sz_i
\sz_j.
\end{equation}
We have not included all possible interaction terms
$V^{\alpha\beta}_{ij} \hat{\tau}_i^{\alpha} \hat{\tau}_j^{\beta}$
here, because the algebra then becomes rather messy, but instead
just all the terms representing different kinds of interaction: the
longitudinal and transverse diagonal couplings $V^{\parallel}_{ij}$
and $V^{\perp}_{ij}$, and a representative non-diagonal $\chi_{ij}$.

\begin{figure}[ht]
\begin{center}
\scalebox{0.7}{\includegraphics{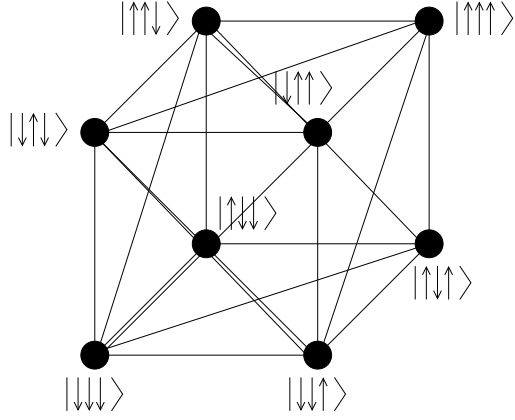}}
\end{center}
\caption{Graph for the quantum walk given by the Hamiltonian
(\ref{eq::HamUQC}). The nodes are labelled as in figure
\ref{fig::cubic-lattice}. The diagonal edges correspond to the
two-qubit terms in the Hamiltonian, while the self-loops come from
the $\sz_{i}$ terms. }
 \label{fig::cubic-lattice-for-UQC}
\end{figure}

It is intuitively useful, before giving the general results, to
first consider just three qubits. Using the binary expansion
encoding, where the state $\ket{k}$ represents the $k^{th}$ node on
some graph, we have
\begin{eqnarray*}
\hat{H} &= & \textbf{[} (\chi_{21} + \chi_{31} +
\Delta_1)\ketb{0}{4} + (\chi_{21} - \chi_{31} + \Delta_1)\ketb{1}{5}
+ (\chi_{31} - \chi_{21} + \Delta_1) \ketb{2}{6} + (\Delta_1 -
\chi_{21} -
\chi_{31}) \ketb{3}{7} \\
&& +  (\chi_{12} + \chi_{32} + \Delta_2)\ketb{0}{2} + (\chi_{12} -
\chi_{32}+ \Delta_2)\ketb{1}{3} + (\chi_{32} - \chi_{12} +
\Delta_2)\ketb{4}{6} + (\Delta_2 - \chi_{32} -
\chi_{12})\ketb{5}{7} \\
&& + (\chi_{13} + \chi_{23} + \Delta_3)\ketb{0}{1} + (\chi_{13} -
\chi_{23} +\Delta_3)\ketb{2}{3} + (\chi_{23} - \chi_{13}
+\Delta_3)\ketb{4}{5} +
(\Delta_3 -\chi_{23} - \chi_{13}) \ketb{6}{7} + \textrm{H.c.} \textbf{]} \\
&&\;\;\;\;\;\;\; +  \textbf{[} V^{\perp}_{12} \left( \ketb{0}{6} +
\ketb{1}{7} + \ketb{2}{4} + \ketb{3}{5}\right) + V^{\perp}_{23}
\left(\ketb{0}{3} + \ketb{1}{2} + \ketb{4}{7} + \ketb{5}{6}\right)
\\ && \;\;\;\;\;\;\; +  V^{\perp}_{13} \left(\ketb{0}{5} + \ketb{1}{4} + \ketb{2}{7}
+ \ketb{3}{6}\right) + \textrm{H.c.} \textbf{]} \\
&& + \textbf{[} (\epsilon_1 + \epsilon_2 + \epsilon_3 +
V^{\parallel}_{12}+ V^{\parallel}_{13} + V^{\parallel}_{23}
)\ketb{0}{0} + (V^{\parallel}_{12}+ V^{\parallel}_{13} +
V^{\parallel}_{23} - \epsilon_1 - \epsilon_2 - \epsilon_3 )\ketb{7}{7}\\
&& + (\epsilon_1 + \epsilon_2 - \epsilon_3 + V^{\parallel}_{12}-
V^{\parallel}_{13} - V^{\parallel}_{23} )\ketb{1}{1} + (
V^{\parallel}_{12}- V^{\parallel}_{13} - V^{\parallel}_{23}
-\epsilon_1 - \epsilon_2 +
\epsilon_3 )\ketb{6}{6} \\
&&(\epsilon_1 - \epsilon_2 + \epsilon_3  - V^{\parallel}_{12}+
V^{\parallel}_{13} - V^{\parallel}_{23} )\ketb{2}{2} + (-\epsilon_1
+ \epsilon_2 - \epsilon_3  - V^{\parallel}_{12}+ V^{\parallel}_{13}
-
V^{\parallel}_{23} ) \ketb{5}{5}\\
&&  + (\epsilon_1 - \epsilon_2 - \epsilon_3 - V^{\parallel}_{12} -
V^{\parallel}_{13} + V^{\parallel}_{23})\ketb{3}{3} + (-\epsilon_1 +
\epsilon_2+ \epsilon_3 - V^{\parallel}_{12} - V^{\parallel}_{13} +
V^{\parallel}_{23})\ketb{4}{4} \textbf{]}.
 \end{eqnarray*}
which is a quantum walk over a cubic lattice, with the addition of
the diagonal connections, on the faces, as well as on-site
potentials, as shown in figure \ref{fig::cubic-lattice-for-UQC}.
If we generalise now to an $N$-qubit Hamiltonian of the form above,
we have a quantum walk on a hypercube, with the addition of
next-nearest neighbour connections, where the nodes are encoded as
described earlier for the hypercube. We can re-express the
Hamiltonian in the general quantum walk form as
\begin{equation}
\hat{H} = -\sum_{\langle ij\rangle}
\Delta_{ij}\left(\hat{c}_{i}^{\dagger}\hat{c}_j
+\hat{c}_{i}\hat{c}_j^{\dagger} \right) + \sum_{j=0}^{2^N}
\epsilon_j \hat{c}_j^{\dagger}\hat{c}_j,
\end{equation}
where the coefficients are defined as follows. Consider the binary
representation of each of the nodes, i.e. $i \equiv i_1 i_2 \ldots
i_N$, $j = j_1 j_2 \ldots j_N$ where $i_a, j_b = 0,1$ corresponding
to spin-up and spin-down in the qubit representation. Then, for $1
\leq a,b \leq N$,
\begin{equation}
\Delta_{ij} = \left\{\begin{array}{ccc} \Delta_a + \sum_c (-1)^{j_c}
\chi_{ca} & \textrm{if } i_a \neq j_a \,\, \textrm{and} \,\,
i_b = j_b \,\, \forall b\neq a \\
V^{\perp}_{ab} &\quad \textrm{if } i_a \neq j_a \,\, \textrm{and}
\,\, i_b \neq j_b  \,\, \textrm{and} \,\,
j_c = i_c \,\, \forall c\neq a,b \\
0 & \textrm{otherwise} \end{array} \right.
\end{equation}
and
\begin{equation}
\epsilon_j = \sum_{a=1}^{N} (-1)^{j_a} \epsilon_a + \sum_{a,b}
(-1)^{j_a + j_b} V^{\parallel}_{ab}.
\end{equation}
The only aspect of these expressions that is not immediately obvious
is the signs.

\subsection{Dynamic Qubit systems mapped to quantum walks}
 \label{sec:dynQB-QW}

We now consider a universal gate set in which we allow
time-dependence in all the couplings. Again we do not consider the
most general case because the results are too messy, but instead
take a special case in which the qubit Hamiltonian has the form
\begin{equation}\label{eq::HamUQC}
\hat{H} = \sum_{j=1}^{N} \left(\epsilon_j(t) \sz_{j} -
\Delta_j(t)\sx_{j}\right) - \sum_{i,j}V^{\perp}_{ij}(t) \sx_i\sx_j,
\end{equation}
where we have complete control over all parameters in the
Hamiltonian, which are time-dependent. This is a rather idealised
case, but will suffice for our demonstration. If every qubit is
`connected', such that there are sufficient coupling terms between
qubits allowing entanglement between all, then the Hamiltonian is universal for quantum computation. The two
single qubit terms allow any single-qubit unitary to be implemented,
then all that is needed is a two-qubit entangling operation
\cite{NBD+02}, as provided by the $XX$ coupling.

From this Hamiltonian, a quantum circuit will correspond to a pulse
sequence, describing applications of different terms in the
Hamiltonian. The fundamental gate set consists firstly of arbitrary
$x$ and $z$ rotations (on the Bloch sphere) for each qubit, denoted
\begin{equation}
R_x(\gamma) = \exp(-i\gamma \sx/2)\! , \quad R_z(\theta) =
\exp(-i\theta \sz /2),
\end{equation}
which can be combined to describe any single qubit unitary operation
$\hat{U}$, via
\begin{equation}
\hat{U} = e^{i\alpha}R_z(\theta)R_x(\gamma)R_z(\xi),
\end{equation}
for some global phase $\alpha$. As well we have the two-qubit
unitaries described by
\begin{equation}
V^{\perp}_{ij}(\chi) = \exp(i\chi \sx_{i}\sx_{j}),
\end{equation}
between qubits $i,j$. We will construct circuits in terms of these
fundamental gates, then convert the relevant pulse sequence into a
quantum walk.

The canonical universal gate set consists of single-qubit unitaries
and the controlled-NOT, ({\sc cnot}) operation. Using a method from Ref.
\cite{BDD+02}, we show below a circuit which is equivalent to {\sc
not} made up gates from our fundamental set;
\[
\Qcircuit @C=0.75em @R=1.0em {
 &\ctrl{2}& \qw & & & \gate{Z_{\frac{-\pi}{2}}} &
 \gate{X_{\frac{\pi}{2}}}&\gate{Z_{\frac{\pi}{2}}} &
 \multigate{2}{W}& \gate{Z_{\frac{-\pi}{2}}} &
 \gate{X_{\frac{\pi}{2}}}&\gate{Z_{\pi}}&\qw \\
& & & \equiv  & & & & & & & & & \\
 & \targ &\qw & & &\qw &\qw &\qw &\ghost{W} &\qw &
 \gate{X_{\frac{-\pi}{2}}} &\qw &\qw}\]
  \label{fig::qcir-cnot}
 For compactness of
notation, we set $R_x(\theta) \equiv X_\theta$ and
$R_z(\xi)\equiv Z_\xi$, and $W = V^{\perp}(\pi/4)$.
The circuit in terms of the fundamental gates easily becomes a pulse
sequence by interpreting the angles as times of application for
corresponding terms in the Hamiltonian. Applying $R_x(\gamma)$ on
the second qubit corresponds to switching on $\Delta_2$ for a time
$T$ such that $T = -\gamma/2\Delta_2$. When $\gamma$ is positive, we
simple replace this with the angle $\gamma' = \gamma - 2\pi$, which
gives an equivalent rotation. Similarly, for $R_z(\theta)$ on the
third qubit, $T = \theta/2\epsilon_2$, and for $V^{\perp}(\chi)$ on
the third and fourth qubits, we switch $V^{\perp}_{34}$ on for a
time $T =\chi/V^{\perp}_{34}$.\\

\begin{figure}[h!]
\begin{center}
\large
\scalebox{0.5}{ \includegraphics{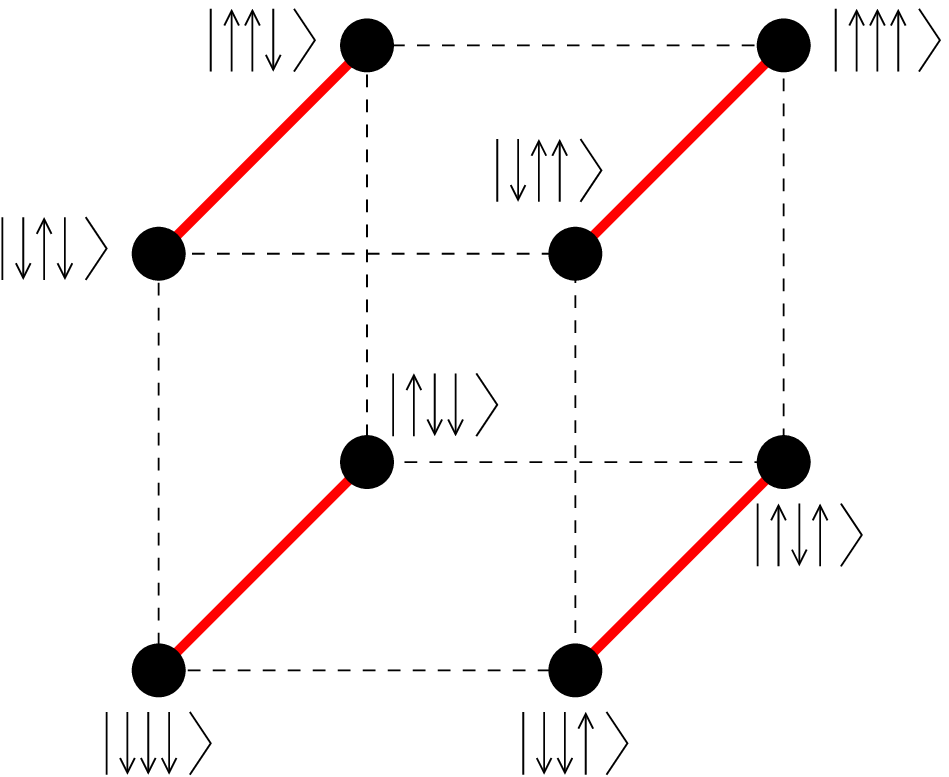}}
\scalebox{0.5}{\includegraphics{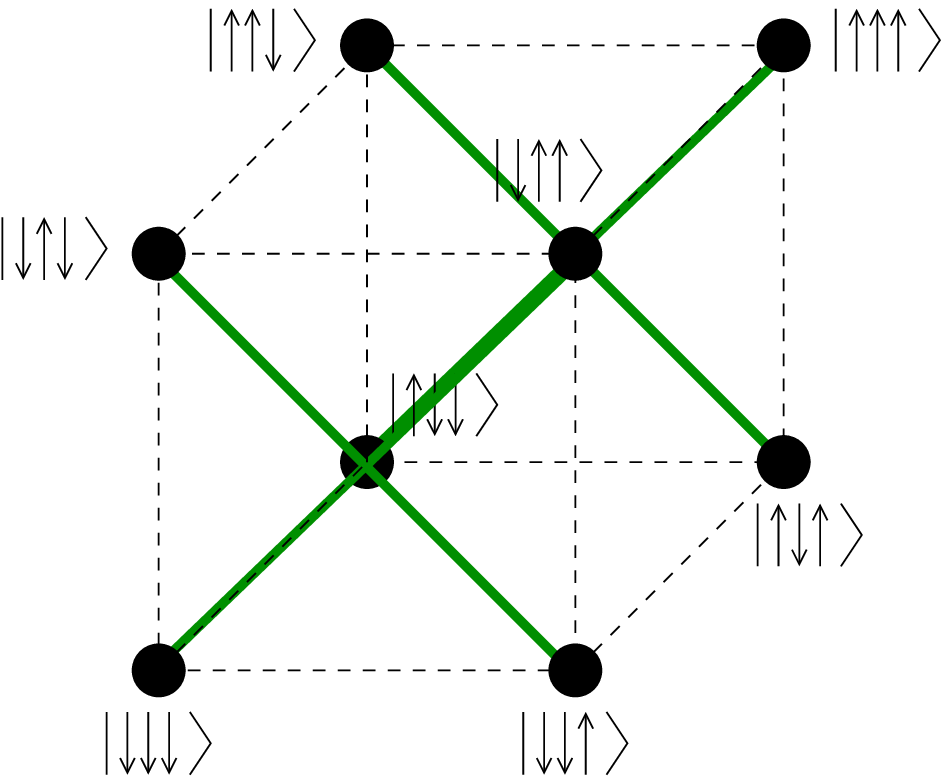}}
\scalebox{0.5}{ \includegraphics{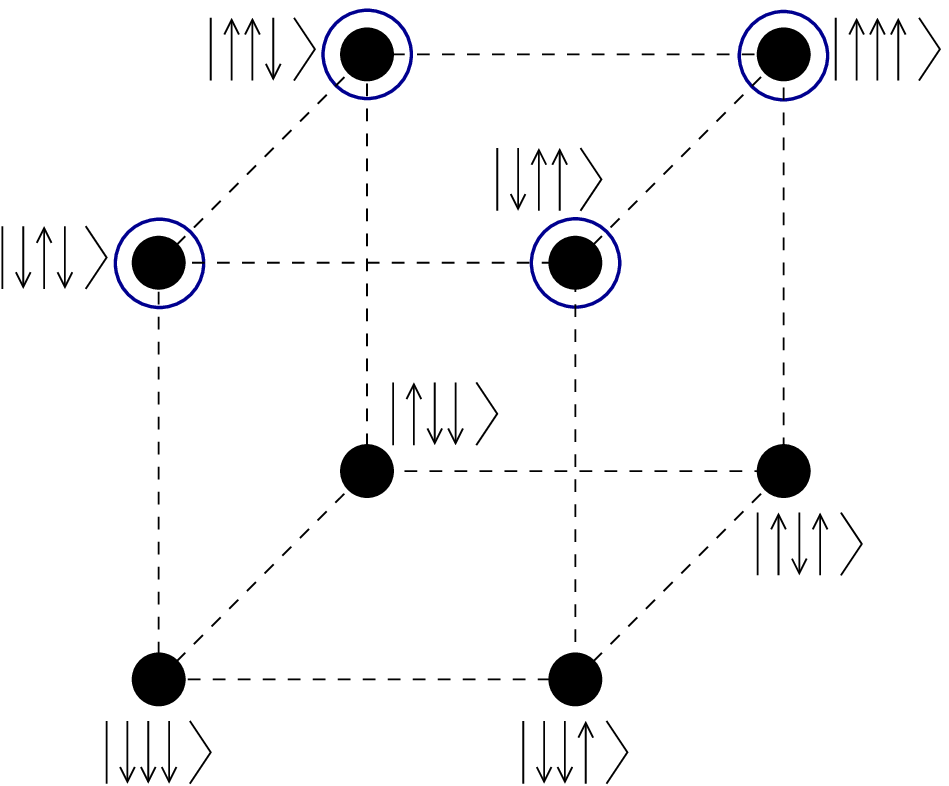}}\\
$$  R_x^{(1)}(\gamma) = \exp(-i\gamma \sx_{1} / 2), \quad\quad V^{\perp}_{23}
(\chi) = \exp(i\chi \sx_{2}\sx_{3}), \quad\quad R_z^{(2)}(\theta) =
\exp(-i\theta \sz_{2} /2)$$
 \end{center}
 \caption{Fundamental gates as variants of a quantum walk on the
hypercube.}\label{fig::gate-qw}
 \end{figure}

We can interpret each fundamental gate in terms of a quantum walk on
graph whose nodes are arranged on the hypercube with the specific
gate determining the edges (see figure \ref{fig::gate-qw}).

Imagine the $2^N$ nodes of a quantum walk arranged on a hypercube.
An $R_x^{(k)}(\gamma)$ pulse switches on connections along edges --
figure \ref{fig::gate-qw}(1) -- in a direction given by the qubit
acted upon. We then have a quantum walk on this restricted
hypercube, for a time corresponding the angle $\gamma$.

Similarly, a $V^{\perp}_{jk}(\chi)$ pulse `switches on' connections
along the diagonals of faces determined by the qubits acted upon,
resulting in a different restricted quantum walk, for a time
corresponding to $\chi$ (figure \ref{fig::gate-qw}(2)).

On the other hand, a $R_z^{(j)}(\theta)$ pulse does not connect any
nodes, but rather applies a relative phase to half of the nodes,
i.e. \begin{equation} R_z(\theta)(a\ket{0}+b\ket{1}) = e^{-i\theta}
(a\ket{0} + be^{i2\theta}\ket{1}).
\end{equation}
This relative phase is applied to the nodes on a `face' of the
hypercube, dependent upon the qubit acted upon (see figure
\ref{fig::gate-qw}(3)). A quantum computation will correspond to a
series of these pulses, of varying time -- the analogous quantum
walk will be over a hypercube with time-dependent edges. As an
example, we consider the quantum Fourier transform (QFT), the
essential element of Shor's factoring algorithm.

The QFT on an orthonormal basis $\ket{0},\ket{1},\ldots,\ket{N-1}$
is defined by the linear operator,
\begin{equation}
\ket{j} \rightarrow \frac{1}{\sqrt{N}} \sum_{k=0}^{N-1} e^{i2\pi
jk/N}\ket{N-1},
\end{equation}
which on an arbitrary state acts as
\begin{equation}
\sum_{j=0}^{N-1} x_j \ket{j} \rightarrow \sum_{k=0}^{N-1} y_k \ket{k},
\end{equation}
where
\begin{equation}
y_k = \frac{1}{\sqrt{N}} \sum_{j=0}^{N-1} x_j e^{i2\pi jk/N}\ket{N-1},
\end{equation}
is the (classical) discrete Fourier transform of the amplitudes
$x_j$. This transformation is unitary, so can implemented on a
quantum computer.

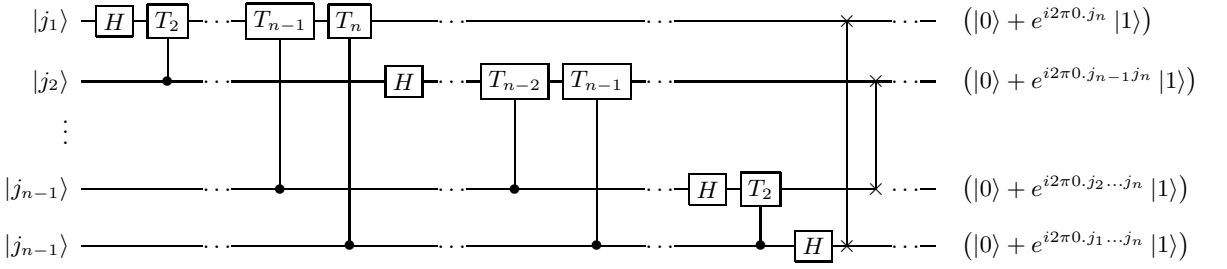
\begin{figure}
\begin{center}
\[
\Qcircuit @C=0.60em @ R=1.0em {
 \lstick{\ket{j_1}} & \gate{H}&\gate{T_2} & \qw & \ldots & & \gate{T_{n-1}} &
 \gate{T_n}& \qw&\qw & \ldots & & \qw&\qw&\qw& \ldots & &\qw &\qw &\qw &
 \qswap &\qw &\qw & \qw & \ldots & &\qw & \rstick{\left( \ket{0} +
 e^{i2\pi 0.j_n}\ket{1} \right)}\\
\lstick{\ket{j_2}} & \qw & \ctrl{-1} &\qw & \ldots & &\qw & \qw
&\gate{H}& \qw & \ldots & &\gate{T_{n-2}} & \gate{T_{n-1}} & \qw &
\ldots & &\qw &\qw &\qw &\qwx \qw & \qw & \qswap&  \qw & \ldots &
&\qw &\rstick{\left( \ket{0} +
e^{i2\pi 0.j_{n-1}j_n}\ket{1} \right)} \\
\lstick{\vdots} & & & & & & & & & & & & & & & & & & & & \qwx &  &\qwx\\
 & & & & & & & & & & & & & & & & & & & & \qwx & & \qwx\\
\lstick{\ket{j_{n-1}}} & \qw &\qw &\qw  & \ldots & & \ctrl{-4} &
\qw& \qw& \qw & \ldots &  &\ctrl{-3} & \qw & \qw & \ldots & &
\gate{H}& \gate{T_2} & \qw  & \qw \qwx & \qw & \qswap \qwx &  &
\ldots &&\qw &
\rstick{\left( \ket{0} + e^{i2\pi 0.j_2 \ldots j_n}\ket{1} \right)}\\
\lstick{\ket{j_{n-1}}} & \qw &\qw &\qw  & \ldots & & \qw &
\ctrl{-5}& \qw& \qw & \ldots & & \qw & \ctrl{-4}&\qw & \ldots & &
\qw &\ctrl{-1} & \gate{H} & \qswap \qwx  & \qw &\qw &\qw & \ldots &
&\qw& \rstick{\left( \ket{0} + e^{i2\pi 0.j_1\ldots j_n}\ket{1}
\right)}}\]
 \end{center}
 \caption{Quantum circuit for the quantum Fourier transform. At the
end are $n/2$ {\sc swap} gates, reordering the qubits.}
  \label{fig::QFT-gen}
 \end{figure}

Following the prescription from \cite{NC00}, to perform the QFT on a
qubit quantum computer we let $N=2^n$, and the basis
$\ket{0},\ldots,\ket{N-1}$ be the computation basis for $n$-qubits.
Each $j$ is expressed in terms of it's binary representation, $j
\equiv j_1j_2\ldots j_{n}$ -- explicitly $j = j_12^{n-1} +
j_22^{n-2} + \ldots + j_{n}2^0$. We use the notation $0.j_k
j_{k+1}\ldots j_{l}$ to represent the \emph{binary fraction} $j_k/2
+ j_{k+1}/4 + \ldots + j_{l}/2^{l-k+1}$. This allows us two write
the action of the QFT in a useful product representation \cite{NC00},
\begin{equation}
\ket{j_1\ldots j_n} \rightarrow \frac{1}{2^{n/2}} \left( \ket{0}
+e^{i2\pi 0.j_n}\ket{1} \right)\left( \ket{0} +
e^{i2\pi0.j_{n-1}j_n}\ket{1} \right)\ldots \left( \ket{0} +
e^{i2\pi0.j_1\ldots j_n}\ket{1} \right).
\end{equation}

Based on this representation, an efficient circuit, shown in figure
\ref{fig::QFT-gen}, for the QFT is constructed \cite{NC00}. This
circuit utilises the Hadamard gate, $H$, {\sc swap} gates, and
controlled-$R_k$ gates, where
\begin{equation}
T_k = \left[\begin{array}{cc} 1 & 0 \\ 0  & e^{i2\pi/2^k}
\end{array}\right].
\end{equation}

\begin{figure}[ht]
\begin{center}
\[
\Qcircuit @C=0.55em @R=0.9em {
& \qswap&\qw & & &\gate{Z_{\frac{-\pi}{2}}} & \gate{X_{\frac{\pi}{2}}} &
\gate{Z_{\frac{\pi}{2}}}&\multigate{2}{V\left(\frac{\pi}{4}\right)}&\gate{Z_{\frac{-\pi}{2}}}&
\gate{X_{\frac{-\pi}{2}}} &\multigate{2}{V\left(\frac{\pi}{4}\right)}
&\gate{Z_{\frac{\pi}{2}}} &\multigate{2}{V\left(\frac{\pi}{4}\right)}&\qw \\
&\qwx& & \equiv & \\
& \qswap \qwx&\qw & & &\gate{Z_{\frac{-\pi}{2}}} &
\gate{X_{\frac{\pi}{2}}} & \gate{Z_{\frac{\pi}{2}}}
&\ghost{V\left(\frac{\pi}{4}\right)}& \gate{Z_{\frac{-\pi}{2}}}&
\gate{X_{\frac{-\pi}{2}}}
&\ghost{V\left(\frac{\pi}{4}\right)}&\gate{Z_{\frac{\pi}{2}}} &
\ghost{V\left(\frac{\pi}{4}\right)}&\qw }\]
 \caption{The {\sc swap}
gate as a pulse sequence using our fundamental
gates.}
 \label{fig::swap}
 \end{center}
 \end{figure}
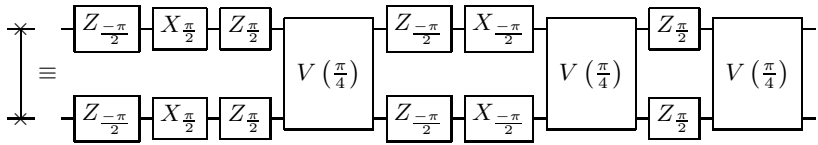

We can rewrite this circuit in terms of our fundamental gate set, to
derive a corresponding pulse sequence. A controlled-$T_k$ gate is
given in figure \ref{fig::CRk}, while the {\sc swap} gate is shown
in figure \ref{fig::swap}.
\begin{figure}[ht]
\begin{center}
\[
\Qcircuit @C=0.60em @R=1.0em {
 &\ctrl{2} &\qw & &\gate{Z_{\frac{-\pi}{2}}} & \gate{X_{\frac{\pi}{2}}}
 &\gate{Z_{\frac{\pi}{2}}} &\multigate{2}{V\left(\frac{\pi}{2^{k+1}}\right)} &
\gate{Z_{\frac{-\pi}{2}}} & \gate{X_{\frac{-\pi}{2}}}
&\gate{Z_{\frac{\pi(2^{k-1}+1)}{2^k}}} & \qw\\
& & & \equiv & \\
& \gate{R_k} &\qw & & \gate{Z_{\frac{-\pi}{2}}}
&\gate{X_{\frac{\pi}{2}}} & \gate{Z_{\frac{\pi}{2}}}
&\ghost{V\left(\frac{\pi}{2^{k+1}}\right)} &
\gate{Z_{\frac{-\pi}{2}}} & \gate{X_{\frac{-\pi}{2}}}
&\gate{Z_{\frac{\pi(2^{k-1}+1)}{2^k}}} &\qw }\]
 \caption{The controlled-$R_k$ gate in terms of the fundamental gate
set. The pulse sequence can be read directly from the
circuit.}
 \label{fig::CRk}
 \end{center}
 \end{figure}
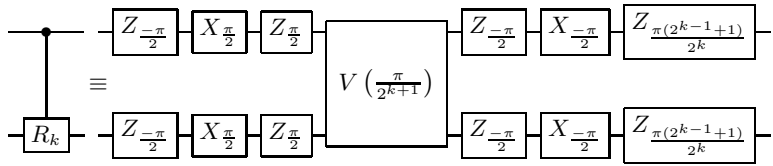

By combining these circuits we construct the QFT circuit in terms of
our fundamental gates set. This circuit can be interpreted as a
pulse sequences, the duration of the pulses corresponding to the
angles characterising the different gates.

For the above example we have assumed complete control over all
parameters in the Hamiltonian, with the ability to switch all on or
off. In physical systems, this is almost surely not the case. For
example, interactions may be constant, with the single qubit terms
controllable. Quantum computation is still possible in this case,
though pulse sequences will be more complicated. An interesting
problem is how circuit complexity varies as further restrictions are
placed on possible controls. The problem of constructing efficient
circuits in general is a very open and active area of research \cite{NDGD06}; when
decoherence is included in the operation of the gates, this becomes
even more interesting -- circuits would be designed to minimise
decoherence, as opposed to complexity. Naively, one would expect
less gates to mean shorter running time and lessening the effects of
decoherence. A detailed study may demonstrate that this is not the
case.


\section{Concluding Remarks}

 \label{sec:Conc}


In this paper we have formulated quantum walks in a Hamiltonian
framework, and explored the mappings that exist between various
quantum walk systems and systems of gates and qubits. The
Hamiltonian formulation possesses considerable advantages. We have
seen that it allows a unified treatment of continuous time and
discrete time walks, for both simple and composite quantum walk
systems. It is also necessary if one wishes to make the link to
experimental systems. This latter point becomes particularly clear
when one tries to understand decoherence for quantum walkers, for
which it is essential to set up a Hamiltonian or a Lagrangian
description.

In the paper we have concentrated on walks on hypercubes and
hyperlattices. Walks on hypercubes are naturally mapped to systems
of gates or qubits, and we have explored mappings in either
direction. Walks on hyperlattices, on the other hand, can be mapped
to qubit or gate systems, but the mappings are not so obvious -- we
have exhibited them, and thereby shown how one could construct an
experimental $d$-dimensional hyperlattice from a gate system. In the
case of both hypercubes and hyperlattices we have exhibited the
general methods for finding these mappings and their inverses, in
sufficient detail that it should now be clear how to make such
mappings for quantum walks on more general graphs.

The practical use of our methods and results does not become
completely clear until we incorporate the environment into our
Hamiltonian description. As indicated in the introduction, this can
be done in a fairly comprehensive way, by using a general
description of environments in terms of oscillator and or spin
baths. The rather lengthy results once this is done appear in a
companion paper to this one\cite{hines07}. Once this is done it
becomes possible to solve rigourously for the dynamics of quantum
walk systems, without using {\it ad hoc} models with external noise
sources. The results can be pretty surprising, as shown by the
results in ref.\cite{PS06} for one particular example.

Ultimately the main reason for the work in the present paper is that
one can bring the work on quantum walks into contact with
experiment, and design experimental systems able to realise
different kinds of quantum walk. In parallel work we have done this
for both a particular ion trap system, and for a particular
architecture of spin qubits\cite{HSM07}. Only in this way will it be
possible to fully realise the potential offered by quantum walk
theory in the lab (and to test it experimentally!).

\acknowledgements
We would like to thank NSERC, PITP, and PIMS for support, and G.J.
Milburn for useful discussions.



\begin{thebibliography}{99}



\bibitem{crystal}    For an introduction to solid-state physics, see
                     A.W. Ashcroft, N.D. Mermin, "{\it Solid-State
                     Physics}" (Holt, Rinehart,and Winston, 1976)

\bibitem{disorder}   An introduction to quantum dynamics
                     on disorded lattices and graphs is in J.M. Ziman,
                     "{\it Models of Disorder}" (CUP, 1979)

\bibitem{Qmag}       An introduction to the quantum magnetism of lattice spin systems
                     is given by P. Fazekas, "{\it Electron
                     Correlation and Magnetism}" (World Scientific,
                     1999); the topic of spin glasses, and related
                     work in optimisation and neural network theory,
                     is treated in K.H. Fischer, J.A. Hertz, "{\it
                     Spin Glasses} (CUP, 1991), and M Mezard, G. Parisi, M.A. Virasoro,
                     "{\it Spin Glass Theory and Beyond}" (World Scientific, 1987)

\bibitem{atom}       Optical lattice systems are reviewed in I. Bloch, Nature Physics
                     {\bf 1}, 23 (2005), and refs therein.

\bibitem{StatM}      Treatments of statistical mechanics emphasizing methods
                     useful for degrees of freedom on lattices and other graphs
                     include D.C. Mattis, "{\it Statistical
                     Mechanics made simple}" (World Scientific,
                     2003), and L.P. Kadanoff, "{\it Statistical
                     Physics: Statics, Dynamics, and
                     Renormalisation}" (World Scientific, 2000).

\bibitem{kempe03}    J. Kempe, Contemp. Phys. {\bf 44}, 307 (2003).

\bibitem{farhi98}    E. Farhi and S. Gutmann, Phys. Rev. {\bf A58}, 915 (1998).

\bibitem{childs02}   A. M. Childs, E. Farhi, and S. Gutmann,
                       Quant. Inf. Proc. {\bf 1}, 35 (2002).

\bibitem{childs02a}  A. M. Childs, R. Cleve, E. Deotto, E. Farhi,
                       S. Gutmann, and D. A. Spielman, Proc. 35th
                       ACM Symposium on Theory of Computing (STOC 2003),
                       pp. 59-68

\bibitem{ambainis}   S. Aaronson and A. Ambainis, Theory of Computing
                     {\bf 1}(4) 47-79 (2005); A. Ambainis, J. Kempe, and
                     A. Rivosh, Proc. 16th ACM-SIAM SODA, p. 1099-1108 (2005).

\bibitem{ambainis03} A. Ambainis, Intl. J. Quantum Inf. {\bf 1} 407 (2003).

\bibitem{childs04}   A. M. Childs and J. Goldstone, Phys. Rev. {\bf A70},
                       022314 (2004); and Phys. Rev. {\bf A70},
                       042312 (2004).

\bibitem{shenvi03}   N. Shenvi, J. Kempe and  K. B. Whaley, Phys. Rev.
                       {\bf A67}, 052307 (2003).

\bibitem{kempe02}     J. Kempe, Probability Theory and Related Fields,
                      Vol. {\bf 133}(2), 215-235 (2005).



\bibitem{kendon03}   V. Kendon,  Int. J. Quantum Info. {\bf 4},  791-805 (2006).

\bibitem{milburn02}  B. C. Travaglione and G. Milburn, Phys. Rev.
                       {\bf A65}, 032310 (2002); W. D\"{u}r, R. Raussendorf,
                       V. M. Kendon, and H-J. Briegel, Phys Rev {\bf A66},
                       052319 (2002); K. Eckert, J. Mompart, G. Birkl and
                       M. Lewenstein, Phys. Rev {\bf A 72}, 012327 (2005);
                       B. C. Sanders, S. D. Bartlett, B. Tregenna, and
                       P. L. Knight, Phys. Rev. A {\bf 67}, 042305 (2003).

\bibitem{fuji05}     S. Fujiwara, H. Osaki, I. M. Buluta, and S. Hasegawa,
                       Phys. Rev. {\bf A72}, 032329 (2005).

\bibitem{bouwmeester99} D. Bouwmeester, I. Marzoli, G. P. Karman,
                       W. Schleich and J. P. Woerdman, Phys. Rev. A {\bf 61},
                       013410 (1999).

\bibitem{ryan05}     C. A. Ryan, M. Laforest, J. C. Boileau and R. Laflamme,
                       Phys. Rev. A {\bf 72}, 062317 (2005).

\bibitem{feyV63}     R.P. Feynman, F.L. Vernon, Ann. Phys {\bf 24}, 118 (1963)

\bibitem{ajl84}      A.J. Leggett, Phys. Rev. {\bf B30}, 1208 (1984)

\bibitem{PS00}       N.V. Prokof'ev, P.C.E. Stamp, Rep. Prog. Phys. {\bf 63}, 669 (2000)

\bibitem{PCES06}     P.C.E. Stamp, Studies Hist. Phil. Mod. Phys. {\bf 37},
                       467 (2006)

\bibitem{hines07}    A. Hines, P.C.E. Stamp, to be published.


\bibitem{PS06}       N.V. Prokof'ev, P.C.E. Stamp, Phys. Rev. {\bf
                       A74}, 020102(R) (2006)

\bibitem{feynman81}  R.P. Feynman, Found Phys. {\bf 16}, 507 (1986)

\bibitem{NC00}       M.A. Nielsen and I.L. Chuang, \textit{Quantum Computation and
                     Quantum Information}, Cambridge University Press (2000).

\bibitem{NBD+02}     M.A. Nielsen. M.J. Bremner, J.L. Dodd, A.M. Childs and
                     C.M. Dawson, Phys. Rev. A {\bf 66}, 022317 (2002).

\bibitem{BDD+02}     M.J. Bremner, C.M. Dawson, J.L. Dodd, A. Gilchrist,
                     A.W. Harrow, D. Mortimer, M.A. Nielsen and T.J. Osborne,
                     Phys. Rev. Lett. {\bf 89} 247902 (2002).
                     
\bibitem{NDGD06} M.A. Nielsen, M.R. Dowling, M. Gu and A.C. Doherty, Science 311, 1133 (2006).


\bibitem{HSM07}      A Hines, G Milburn, and PCE Stamp, to be
                     published.


\end{thebibliography}
\end{document}